# Optimization and effect of UV-ozone exposure of electron transport layer on the efficiency of the dye-sensitized solar cells

Chandan Dawo[a], Mohammad Adil Afroz[b], Parameswar Krishnan Iyer[b,c] and Harsh Chaturvedi*[a]

[a]Centre for Energy, Indian Institute of Technology Guwahati, Guwahati 781039, Assam, India

[b]Department of Chemistry, Indian Institute of Technology Guwahati, Guwahati 781039, Assam, India

[c]Centre for Nanotechnology, Indian Institute of Technology Guwahati, Guwahati 781039, Assam, India

**Corresponding Author:**    *Harsh Chaturvedi, email: harshc@iitg.ac.in

**ABSTRACT:**

The surface states of the active $TiO_2$ layer is crucial while fabricating an efficient solar cell. This work experimentally analyses the effect of exposing $TiO_2$ based electron transport layer (ETL) to the ultraviolet-ozone (UV-$O_3$) and optimizes the exposure time for improving power conversion efficiency (PCE) of fabricated dye-sensitized solar cells (DSSCs). These results demonstrate that the performance of DSSCs can be improved significantly by UV-$O_3$ exposure of sintered $TiO_2$ photoanode surface, with the duration of exposure being a critical parameter. Fabricated devices show 33.01 % increase in PCE for the optimum exposure. Nevertheless, overexposure of the sample beyond the optimum time decreases the efficiency of the fabricated solar cells. The device with optimum exposure exhibits the highest PCE of 8.34% with short circuit current density ($J$sc) of 15.15 mA/cm$^2$, open circuit voltage ($V$oc) of 756 mV and Fill factor (FF) of 71.10%. This increase in efficiency is attributed to the enhanced crystallization and reduction in the organic contaminants C-C/C-H from 57.90 to 52.40% as shown by the X-ray diffraction (XRD) and X-ray photoelectron spectroscopy (XPS), respectively. The XPS result further indicates an increase in oxygen vacancy from 12.40 to 23.40% for O 1s state and from 9.30 to 14.30% for Ti 2p state of $Ti^{3+}$. Results from the Atomic Force Microscope (AFM) also confirms the minimized surface roughness of 16.36 nm for the optimally exposed $TiO_2$ film, and increase in hydrophilicity leading to improved efficiency of the solar cells which were optimally exposed to UV-$O_3$.

## 1. INTRODUCTION

Grätzel and O'Reagan developed dye-sensitized solar cells (DSSCs) in 1991 with mesoporous $TiO_2$ film as photoanode [1]. Since then, DSSCs have drawn the attention of the scientific community because of its low materials cost, easy fabrication process, lightweight, flexibility, tunable optical properties, and raw materials availability. DSSCs are generally composed of mesoporous $TiO_2$ photoanode to uptake dye photosensitizer, liquid electrolytes ($I^-/I_3^-$), and platinum (Pt) counter electrode. So far, ≈13% efficiency has been reported for DSSCs based on $TiO_2$ photoanode [2,3].

The most preferred photoanode material for DSSC is nanocrystalline $TiO_2$ due to its electro-optical characteristics i.e., favourable valence band and conduction band position for dye and FTO, chemical stability, high refractive index (2.45), cost-effectiveness, strong oxidizing power, mesoporous nature and low toxicity [4-6]. $TiO_2$ based photoanode plays multiple roles for efficient charge generation, charge separation, and charge transfer from dye molecule to transparent conducting oxide (TCO) substrate. The conversion of light to electricity consists of multiple dynamic processes in DSSCs. Several undesired recombination processes are also known to occur at the photoanode, significantly reducing the PCE. Minimizing such unwanted processes at the active layer of $TiO_2$ is required for the efficient performance of the DSSCs. Hence, the surface states of the active layer of $TiO_2$ plays an important role in the fabrication of an efficient solar cell [7-11].

Several modifications of the $TiO_2$ photoanode have been reported to increase the efficiency of DSSCs, such as compact or blocking layer [12-14], scattering layer [15], atomic doping [16, 17], surface treatment [18], etc. The surface treatment of $TiO_2$ electron transport layer (ETL) has become an essential step for improving the performance of DSSCs. Surface treatment of the photoanode affects dye loading capacity, electron transport, and electron recombination processes. $TiCl_4$ treatment on the $TiO_2$ ETL has been the most commonly applied method to increase efficiency of DSSCs [19-21]. However, TiCl$_4$ is not stable at room temperature as it reacts with moisture present in the air to produce the harmful hydrochloric acid [22]. A simple post-treatment of TiO$_2$ photoanode by urea solution with a concentration of 1 g/mol showed improvement in the efficiency of solar cells [23]. A comparative study of oxygen plasma treatment and oxygen ion beam treatment has reported that Oxygen ion beam treatment of DSSCs shows higher efficiency than plasma treatment. As it is much more efficient in removing the number of oxygen vacancies from a TiO$_2$ active film [24]. Deposition of various other layers of Ba(NO$_3$)$_2$, N$_2$O$_6$Sr, and Mg(NO$_3$)$_2$, on TiO$_2$ electrode followed by an O$_2$ plasma treatment, also shows significant improvement in the photo-conversion efficiency of the DSSCs [25]. Oxygen (O$_2$) plasma treatment of TiO$_2$ ETL, remarkably increase in the performance of DSSCs [26]. The drawback of oxygen plasma treatment, however, is its cost, which requires considerable capital investment. Therefore, UV-O$_3$ treatment provides an effective and economical alternative method [27].

In this study, an attempt has been made to study and optimize the effect of UV-O$_3$ exposure on a uniform TiO$_2$ ETL leading to DSSCs with higher efficiency. The samples were exposed to UV-O$_3$ for different duration of times (0, 5, 10, 15, and 30 min), under ambient conditions, and keeping other parameters constant. The results demonstrate that by optimizing the exposure time, we can effectively control the stoichiometry of these samples. This results in an optimized condition affecting optical absorption, wettability, crystallinity, surface morphology, electrical conductivity etc., thereby improving the overall carrier dynamics of the prepared TiO$_2$ film. Consequently, at the optimum condition of 10 min exposure, the PCE of DSSCs shows significant improvement to 8.34% as compared to the untreated device with a PCE of 6.27 %.

## 2. EXPERIMENTAL

### 2.1 Chemicals and Materials

All chemicals and reagents purchased for the experiments were used as received and obtained from different commercial sources: FTO (Fluorine doped tin oxide) glass- 7 Ω/square sheet resistance (Sigma- Aldrich), TiO$_2$ nanoparticles (Degussa P25, Sigma-Aldrich, 21 nm), TiO$_2$ Paste (Sigma-Aldrich, 22 nm), terpineol (Sigma- Aldrich), tert-butanol (Sigma- Aldrich), acetonitrile (fisher scientific), absolute ethanol (Changshu yangyuan chemical, China) , acetic acid (Merck), ethyl cellulose (Himedia), iodine (Himedia), lithium iodide (Sigma-Aldrich), valeronitrile (sigma-

Aldrich), 4-tert butylpyridine (Sigma-Aldrich), 1-butyl-3-methylimidazolium iodide (Sigma-Aldrich), NaOH (Merck), Chloroplatinic acid hexahydrate ($H_2PtCl_6 \cdot 6H_2O$, Sigma-Aldrich), Anhydrous isopropanol (Sigma-Aldrich).

## 2.2 Preparation of $TiO_2$ Working Electrode

FTO substrates were cleaned using 15 min ultrasonication in Mili-Q water, acetone, and 2-propanol, respectively. Cleaned and hot air dried FTO coated substrate was then exposed to UV-$O_3$ treatment for 15 min to remove any organic contaminants. The mesoporous layer of $TiO_2$ consists of two layers. The $TiO_2$ paste consists of 600 mg P25 powder, 5 mL ethanol, 100 µL acetic acid, 500 µL milli-Q water, 3 gm tarpineol, and 3 gm ethyl cellulose (15 wt. % in ethanol). The mixture was stirred overnight and spin-coated at 5000 rpm for 45 secs on the clean substrate. After drying the first layer of $TiO_2$ at 100 °C for 10 min, the second layer of $TiO_2$ was applied by the doctor blade technique using Sigma paste and baked at 100 °C for 20 min. Samples were then sintered at 500 °C for 30 min to improve the interfacial contact between the deposited $TiO_2$ layer and the FTO substrate. $TiO_2$, which acts as a layer for electronic transport, was exposed to UV-$O_3$ for different time durations (0, 5, 10, 20, and 30 min). A set of 10 samples were prepared for each time interval to confirm the reproducibility of the experiment.

## 2.3 Fabrication of the Dye-sensitized solar cell

The working $TiO_2$ photoanodes with and without UV-$O_3$ treatment were dipped into N719 dye solution (0.3 mM N719 in tert-butyl alcohol and acetonitrile, volume ratio 1:1) and kept for 24 hours to sensitize it at ambient condition. The Pt counter electrode was fabricated on another pre-cleaned FTO substrate by spin coating chloroplatinic acid hexahydrate solution (5 mg $H_2PtCl_6 \cdot 6H_2O$ in 1 ml of anhydrous 2-propanol) at 5000 rpm for 45 secs followed by annealing at 500 °C for 30 min. The solution of $I^-/I_3^-$ electrolyte was made with 0.5M LiI, 0.05M $I_2$, 0.5M 1-Butyl-3-methylimidazolium iodide, 0.1M guanidium thiocyanate, and 0.5M 4-tert-butylpyridine dissolved in the mixture of acetonitrile/valeronitrile (v/v= 85:15) solvent. DSSCs were assembled between dyed $TiO_2$ photoanode and Pt counter electrode using a hot melt spacer (Surlyn film, 60 µm, purchased from ossila). The interspace available between the $TiO_2$ photoanode and the counter electrode was then filled with a liquid electrolyte to complete the fabrication of DSSCs. The active area of DSSCs was 0.09 $cm^2$.

## 2.4 Characterization

X-ray photoelectron spectroscopy (XPS) was used to study the structural and chemical changes of $TiO_2$ film, exposed to the UV-$O_3$ system for a different duration. The absorption spectra of thin-film $TiO_2$ photoelectrodes sensitized with N719 dye was measured with diffused reflectance UV-vis spectrometer (Lambda 950). The $TiO_2$ films with dimensions of 0.5 cm x 1.0 cm and a thickness of 8.5 μm were soaked in 3 mL of 0.1 M NaOH solution in ethanol and DI water (volume ratio of 1:1) to desorbed the dye from the film and absorption spectra was taken by means of UV-vis spectrometer (Carry 100). With the help of an automatic liquid dispenser at room temperature, contact angle measurements were done by KRÜSS Drop Shape Analyser (DSA25). The root mean square (RMS) roughness was computed with atomic force microscopy (Agilent 5500-STM). The phase and crystal structure of the $TiO_2$ ETL films was analysed by X-ray diffraction (XRD) with an operating voltage of 45 KV and 200 mA current using Rigaku SmartLab with Cu-K$\alpha$ radiation ($\lambda$=1.54Å). The $TiO_2$ thin films were scanned from 20° to 70º of 2θ with a scan rate of 5 degrees/min. The electrical characteristics of DSSCs were measured with a Keithley 2400 source meter from -1 to 1V in a step of 10 mV, under AM 1.5 G (100 mW/cm$^2$) illumination, which was provided by a solar light simulator (Newport, Oriel Sol 3A). The light intensity of the solar simulator was calibrated with a standard silicon solar cell of the National Renewable energy laboratory (NREL). Using an electrochemical workstation (CH 680), the electrochemical impedance spectra (EIS) of DSSCs were measured in a bias potential particularly at open circuit voltage under dark conditions, in the frequency ranges of 0.1 Hz to 1 MHz with a small alternating signal of amplitude 10 mV. The impedance parameters of the equivalent circuit were determined by fitting the spectra using EC Lab software.

## 2.5 UV-$O_3$ exposure on $TiO_2$ ETL

UV-$O_3$ exposure was performed using NOVASCAN PSD Pro UV system. The UV-$O_3$ system produces ultraviolent light of two wavelengths, 254 and 185 nm simultaneously with very high photon energy of 470 and 647 kJ/mol, respectively. The bond energy of the organic contaminants C-C, C-H, C-O and O-H are 346, 411, 358 and 459 kJ/mol, respectively. The photon energies irradiated from the UV-$O_3$ system are much higher than the bond energies of contaminants. Therefore, the organic contaminant molecules are dissociated or excited by the irradiation of UV light. The atmospheric oxygen ($O_2$) is irradiated with ultraviolet rays at 185 nm; after absorption, it converts to atomic oxygen (O*) and ozone ($O_3$) (equation 1), whereas $O_3$ decomposes after irradiation with 254 nm UV light (equation 2). Strong oxidizing atomic oxygen (O*) is generated during the process of decomposition or formation of $O_3$. Atomic oxygen and $O_3$ strongly react with organic materials to produce volatile molecules such as $CO_2$, $H_2O$, $N_2$ etc. The samples treatment was performed at normal atmospheric pressure and room temperature [28, 29].

$$h\vartheta(185\ nm) + 2O_2 \rightarrow O_3 + O^* \qquad (1)$$

$$h\vartheta(254\ nm) + O_3 \rightarrow O_2 + O^* \qquad (2)$$

## 3. RESULTS AND DISCUSSION

The structural and chemical changes on the $TiO_2$ ETL were studied by XPS upon UV-$O_3$ exposure. Fig. 1a, b demonstrates the XPS spectra of carbon contaminants present in $TiO_2$ film for 0- and 10-min UV-$O_3$ treatment. The C 1s peak intensity found at 284.8 eV decreased significantly after UV-$O_3$ exposure. This suggests that UV-$O_3$ treatment burns out organic contaminants from the precursor of $TiO_2$ film [30, 31]. The quantitative analysis of carbon atom percent, C-C/C-H peak area decreased from 57.90% to 52.40% after UV-$O_3$ treatment (Table S1). The relative changes in the peak area reveals that chemical environment of $TiO_2$ film changes with UV-$O_3$ exposure. Carbon contamination acts as interfacial trap sites and decreases cell performance. It affects the anchoring of dye molecules on $TiO_2$ film, which results in less quantity of dye adsorbed and therefore produces less photocurrent, fill factor, and efficiency. The O1s peak of $TiO_2$ film without and with UV-$O_3$ treatment was fitted with two Gaussian peaks (Fig. 1c, d). The central O1s peak found at 529.99 eV corresponds to the lattice oxygen of $TiO_2$ ($Ti^{4+}$). The higher binding energy on shoulder peak at 531.3 eV±0.2 eV is attributed to the formation of $Ti^{3+}$ surface state through the creation of oxygen vacancies, commonly written as $Ti_2O_3$. The change of oxidation state from $Ti^{4+}$ to $Ti^{3+}$ occurred due to $O^-$ and $O_2^-$ species introduced on the surface of $TiO_2$ by UV-$O_3$ exposure [32-35]. The ratio of the atom percent calculated from the $Ti_2O_3$ peak area increases from 12.40% (0 min) to 23.4% (10 min) after the UV-$O_3$ treatment (Table S1). Oxygen vacancies generated from the reduction of $Ti^{4+}$ to $Ti^{3+}$ produces electrons, thereby affecting the surface functionality and charge state of the $TiO_2$ ETL. The excess number of electrons generated by UV-$O_3$ treatment helps in efficient charge transport and substantial reduction in electron-hole recombination rate as inferred through impedance measurements. Fig. 1e, f illustrates the XPS spectra of Ti 2p for $TiO_2$ samples. For 0 and 10 min UV-$O_3$ exposure, the $TiO_2$ sample shows peaks at 457.5±0.1 eV, and

463±0.2 eV corresponding to the binding energy level of $Ti^{4+} 2p_{3/2}$ and $Ti^{4+} 2p_{1/2}$, along with the peaks at 456.4 eV ±0.1 eV and 458.5 eV ± 0.2 eV imputed to $Ti^{3+} 2p_{3/2}$ and $Ti^{3+} 2p_{1/2}$ energy level of $Ti^{3+}$ state respectively [18, 31, 36, 37].

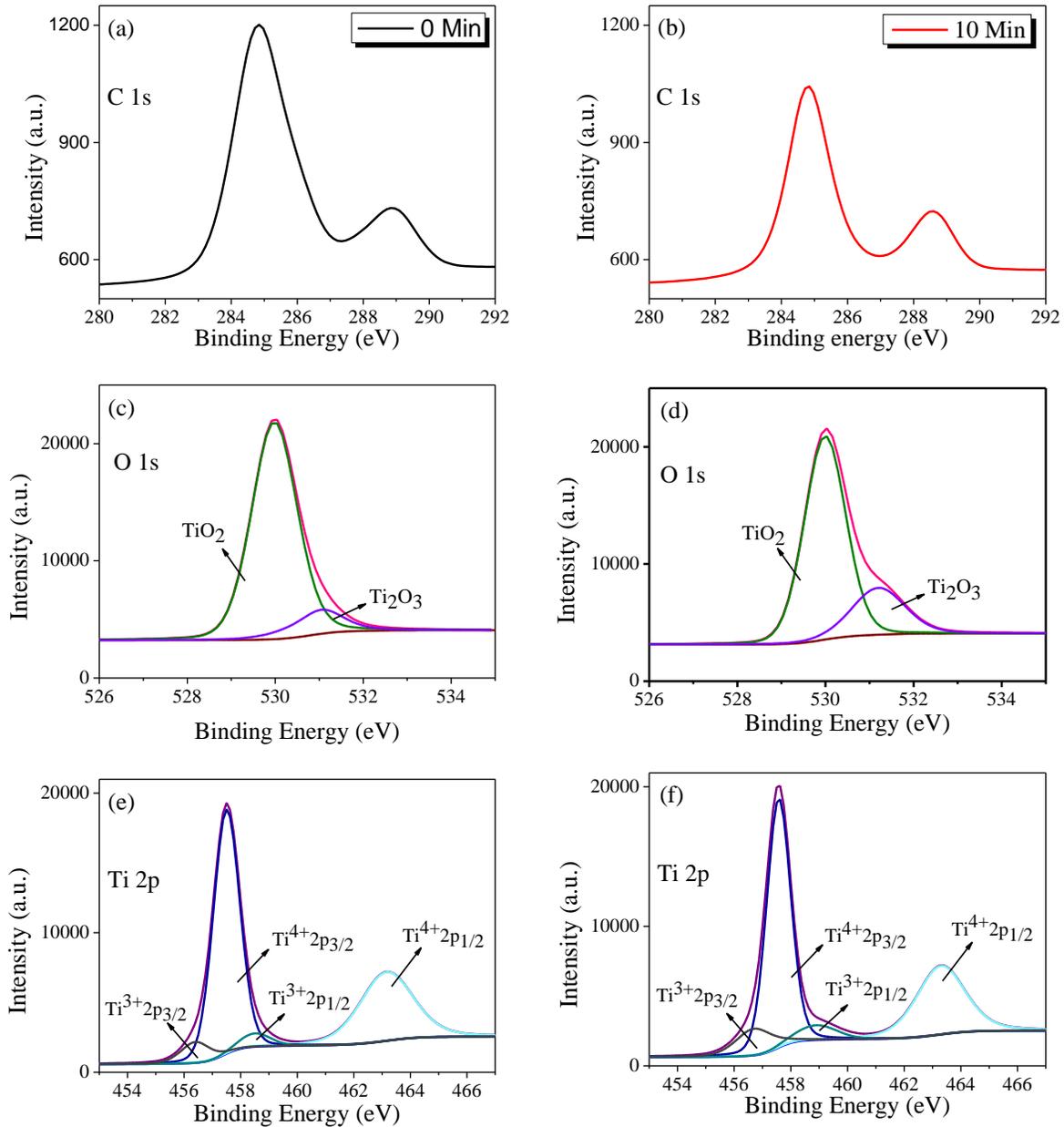

Fig. 1: XPS spectra of $TiO_2$ films on FTO glass substrates : (a, c, and e) with 0 min, and (b, d, and f) with 10 min UV-$O_3$ treatment.

Peak areas of $Ti^{3+}2p_{3/2}$ state increases by 59.60% and $Ti^{3+}2p_{1/2}$ by 44.44% respectively after UV-$O_3$ treatment (Table S1). The difference in these peak areas is attributed to the change in the stoichiometry of the $TiO_2$ film; due to increase in the number of oxygen vacancies by UV-$O_3$ exposure.

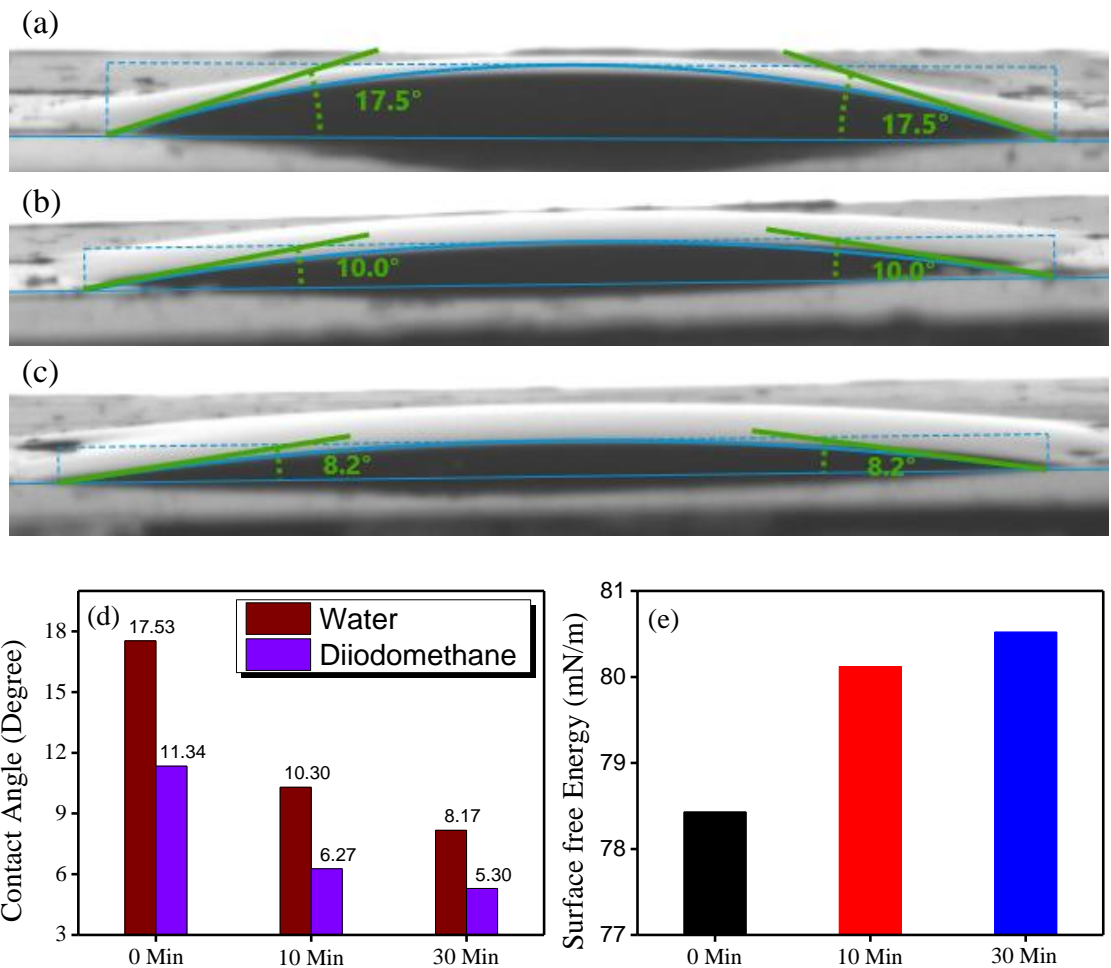

Fig. 2. Contact angle analysis pattern of water on $TiO_2$ films with (a) 0 min (b)10 min and (c) 30 min UV-$O_3$ exposure. Variation of (d) contact angle and (e) Surface free energy by UV-$O_3$ treatment time.

The contact angle of $TiO_2$ active layer films were measured using KRÜSS Drop Shape Analyser instrument at ambient conditions. Deionized (DI) water ($H_2O$) and Diiodo-methane ($CH_2I_2$) were used as the test liquids. The test liquid was placed at different locations on $TiO_2$ active surface. Once the drop achieved metastable equilibrium, the measurement was carried out. Three measurements were taken from each sample, and the mean contact angle value is used to calculate the Surface free energy (SFE). The dispersive and polar components of the surface free energy were computed from the measured contact angles of DI water and Diiodo-methane using KRÜSS advance software (Table S2). The following equation was used to calculate the total SFE (equation 3) [38].

$$\gamma_S = \gamma_S^d + \gamma_S^p \qquad (3)$$

Where $\gamma_S$ is the SFE, $\gamma_S^d$, and $\gamma_S^p$ are the dispersive and polar component of the SFE of the examined $TiO_2$ films.

Fig. 2a, b, c demonstrates the contact angle analysis of $TiO_2$ photoanodes after sintering without and with UV-$O_3$ irradiation. The contact angle reduces from $17.5 \pm 0.5°$ to $8.2 \pm 1°$ after 30 min of UV-$O_3$ exposure as shown in Fig. 2d. Decrease in the contact angle (Fig. 2a, b, c) indicates increase in hydrophilicity and corresponding increase in surface free energy (Fig. 2e). This Increase in surface energy is attributed to the removal of unwanted hydrophobic organic components and the increase of hydrophilic oxygen vacancies in $TiO_2$ film [34, 39-41].

Atomic force microscopy (AFM) was utilized to analyse the surface topography and roughness/sectional profile of $TiO_2$ ETL film exposed to UV-$O_3$ system for different durations (Fig. 3a, b, c,). The root mean square (RMS) roughness decreases by 27.22 % for 10 min exposure as compared to unexposed sample. However, roughness further increases by 19.52 % for 30 min exposure. High roughness factor and irregular pores affect crystallization properties, increase the grain size of the TiO$_2$ film [39, 42]. 10 min UV-O$_3$ exposure has the lowest surface roughness indicating smooth, uniform pores and compact surface films, leading to efficient harvesting of light, better accessibility of dye adsorption, and prevents direct contact between TiO$_2$ and dye molecules thereby resulting in high conversion efficiency. The highest conversion efficiency in perovskite solar cells was also achieved for the sample with lowest RMS value [33].

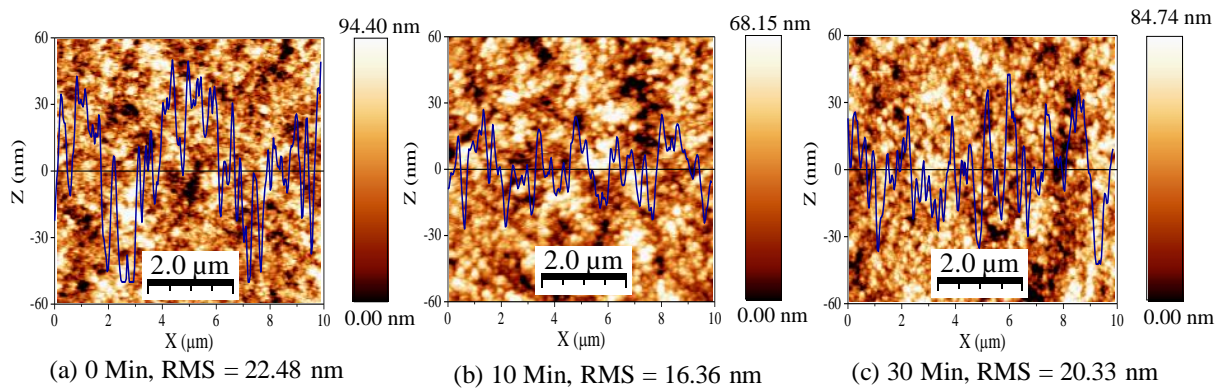

(a) 0 Min, RMS = 22.48 nm  (b) 10 Min, RMS = 16.36 nm  (c) 30 Min, RMS = 20.33 nm

Fig. 3: Two-dimensional AFM images and sectional profile of $TiO_2$ film with (a) 0 min, (b) 10 min, and (c) 30 min UV-$O_3$ exposure.

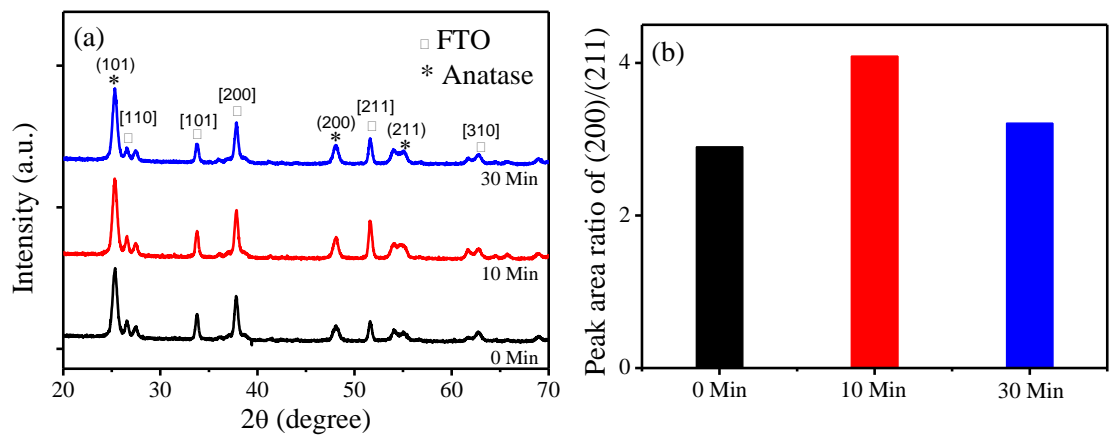

Fig. 4: X-ray diffraction spectra of TiO$_2$ film with (a) 0, 10, and 30 min UV-O$_3$ exposure. (b) Peak area ratio of (200)/ (211) plane.

Fig. 4a displays the XRD pattern of TiO$_2$ film with different UV-O$_3$ treatment time. All samples show diffraction peaks of (101), (200), and (211) planes at corresponding Bragg angles of 25.35°, 48.06°, and 55.00° respectively, and are of anatase phase [8, 43]. The diffraction pattern of TiO$_2$ ETL photoanode with different UV-O$_3$ treatment times remains almost the same, with slight difference in intensity and relative peak area. The highest intensity and relative peak area are observed for 10 min UV-O$_3$ treatment. Changes in the relative peak area are presented in Table S3. Relative area of (211) plane decreased from 10.16% to 6.20% after UV-O$_3$ treatment for 10 min. The highest ratio of 4.09 for (200)/ (211) plane peak area is shown in Fig. 4b. From the relative peak area analysis, it can be concluded that UV-O$_3$ treatment affects the orientation of mesoporous TiO$_2$ ETLs. Favourable atomic arrangement of (101) and (200) orientation leads to improve crystallinity and better interconnection between the TiO$_2$ nanoparticle and FTO.

The changes in the photocatalytic activity were examined using the UV-visible spectrometer. Fig. 5a shows the diffused reflectance UV-Vis absorption spectra of N719 sensitizer anchored on TiO$_2$ photoanode. The photoanode absorption spectra shows two metal-to-ligand charge-transfer (MLCT) peak at 534 nm and 400 nm. Fig. 5b illustrates the absorption spectra of dye desorbing from the TiO$_2$ photoanode. Characteristic changes and blue shifts are observed in absorption pattern for TiO$_2$ loaded with dye. This blue shift is attributed to better deprotonation and less aggregation of the dye on TiO$_2$ surface [3, 9, 44, 45]. Increase in absorption intensity is observed for corresponding exposure to UV-O$_3$ treatment (0 min <10 min< 30 min). This increase in absorption intensity of the photoanode sample exposed to UV-O$_3$ indicates efficient adsorption of the dye molecules on the TiO$_2$ surface due to removal of organic contamination there by providing better surface contact. This increase in dye loading capacity due to UV-O$_3$ exposure, leads to significant improvement in the current density of the fabricated DSSCs.

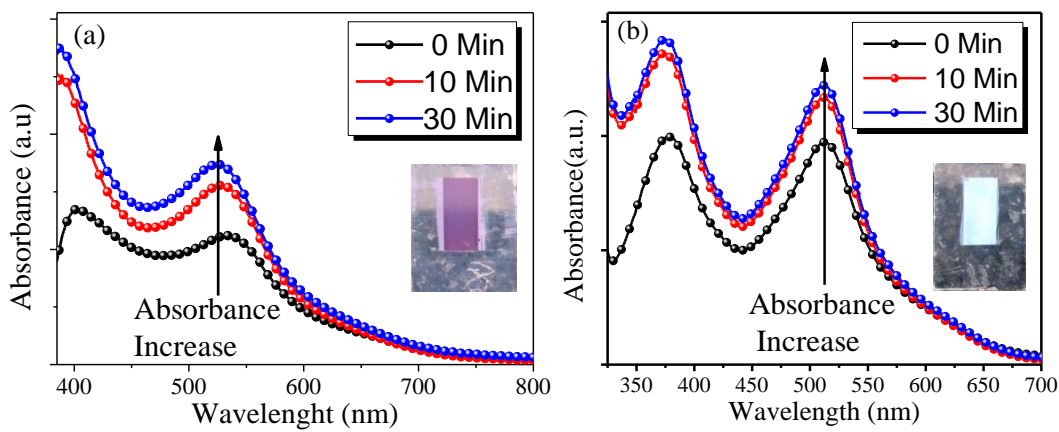

Fig. 5: UV-Visible absorption spectra of N719 dye (a) loaded on TiO$_2$ film, inset shows sensitized film and (b) dye solution desorbed from TiO$_2$ film, inset shows desorbed film.

Fig. 6 shows the Nyquist plots of electrochemical impedance spectra (EIS) for DSSCs along with the equivalent circuit (inset of fig. 6). EIS measurement of DSSCs is a powerful technique to study the charge transfer resistance and recombination reactions taking place at different interfaces in the solar cell i.e., first semicircle in the high-frequency region (from 1-100 kHz) explains the charge transfer resistance of counter electrode/electrolyte interface, second semicircle in the mid-frequency region (from 0.1 Hz-1 kHz) corresponds to the charge transfer resistance at photoanode/electrolyte interface, and in the low frequency region (from 0.1 Hz-0.01 Hz) provides information about diffusion resistance of the electrolytes [44, 46, 47]. The parameters of equivalent circuit as obtained by fitting the data are presented in Table S4. In the high frequency region, the onset point on the real axis of first semicircle corresponds to ohmic series resistance (R$_s$) of the

FTO substrate and TiO$_2$ catalytic film. The value of R$_s$ decreases from 29.98Ω (0 min) to 20.25Ω (30 min) after UV-O$_3$ exposure. This decrease in R$_s$ indicates better interfacial contact of the catalytic film on FTO leading to increase conductivity.

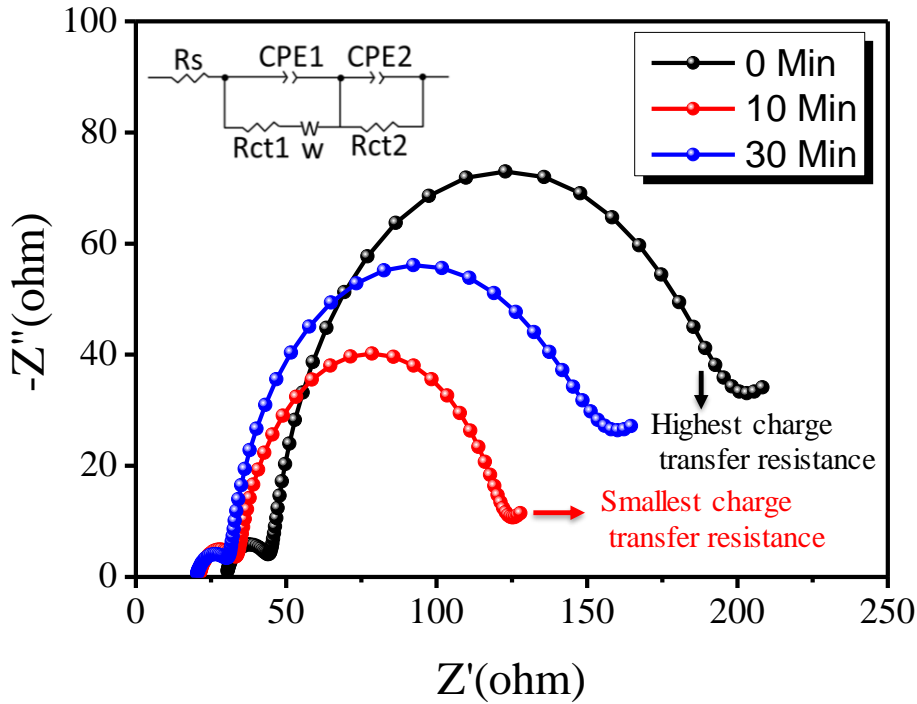

Fig. 6: Nyquist plot of DSSCs under the dark condition for different UV-O$_3$ treatment time.

The smallest semicircle (red colour) in the mid-frequency region for optimum 10 min treatment indicates substantial decrease in the charge transfer resistance, indicating minimum trap state, lower recombination rate, high electro-catalytic activity and faster diffusion of the electrolyte. Large semi-circle (black colour) for untreated device shows considerable charge transfer resistance leading to slower electronic transport, and increase in recombination reaction at the interface. This, results in overall lower performance of the DSSCs.

The electrical conductivity of TiO$_2$ ETL significantly influences the performance of DSSCs. With different UV-O$_3$ irradiation time, the linear scan of dark current (I) vs. voltage (V) curves of TiO$_2$ samples are shown in Fig. 7a. The slope of the I-V curves provides the electrical conductivity of the TiO$_2$ photoanode. This significant increase in I-V slope with increasing UV-O$_3$ exposure time is observed due to improved interfacial contact. UV-O$_3$ exposure of TiO$_2$ films may lead to enhanced conductivity due to the generation of oxygen vacancies [48, 49].

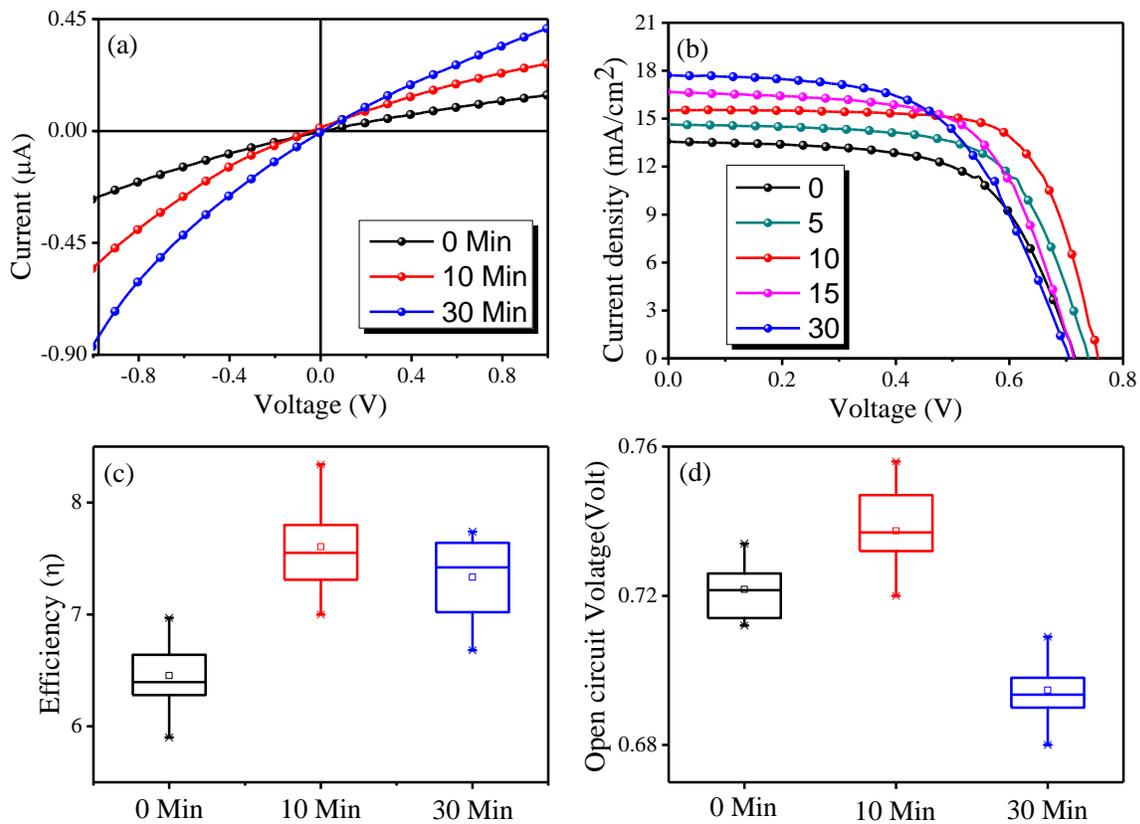

Fig. 7: Effect of UV-O$_3$ treatment on TiO$_2$ ETL (a) Dark linear *I-V* characteristics (b) Light *J-V* curve of DSSCs (c and d) Box chart of efficiency and open circuit voltage.

In order to understand the influence of UV-O$_3$ exposure, the performance of DSSCs *J-V* curve is displayed in Fig 7b. The PCE of DSSCs with different UV-O$_3$ exposure time are shown in Table S5. At 0 min, the photoelectric conversion efficiency (η) of DSSCs is 6.27% ($V_{oc}$=718.40 mV, $J_{sc}$ = 13.02 mA/cm$^2$ and FF = 64.20%). The presence of residual organic contaminants on the TiO$_2$ film acts as recombination centre at the interface of TiO$_2$ / N719 dye/electrolytes thereby critically influencing the collection and transport of electrons, thus, affecting the overall efficiency of the DSSCs [33].

Compared to unexposed TiO$_2$ ETL, UV-O$_3$ exposure shows remarkable improvement in photoelectric conversion efficiency (Fig. 6c). By optimizing the duration of exposure, the best PCE of 8.34% achieved for 10 min with an increase of 33.01% efficiency (η= 6.27 to 8.34%), 16.27% photocurrent density ($J_{sc}$= 13.02 to 15.15 mA/cm$^2$), 5.20% open circuit voltage ($V_{oc}$= 718.40 to 756 mV) and 10.74% fill factor (FF= 64.20 to 71.10%). The enhancement of PCE with UV-O$_3$ exposure may be attributed to fast electron transfer, increase in dye absorption, minimization of surface roughness and removal of organic binders [33-34]. However, the longer duration of UV-$O_3$ exposure time for 30 min gives no further enhancement in the performance of DSSCs. The efficiency decreases by 14.14% (η= 8.34 to 7.16%). This decrease in efficiency is attribute to lower $V_{oc,}$ (Fig. 6d) and FF (Fig. S2) in overexposed surface may be due to the role of oxygen vacancies in facilitating carrier recombination at the $TiO_2$/dye interface [33-35, 48].

## 4 CONCLUSIONS

Herein, the effect of UV-$O_3$ exposure on the efficiency of the DSSCs has been carefully analysed and optimized. The fabricated device with optimized time for 10 min UV-$O_3$ exposure shows the best PCE of 8.34% ($J_{sc}$=15.15 mA/cm$^2$, $V_{oc}$=756.00 mV, FF=71.10%) with good reproducibility. However, any further increase in exposure time leads to decrease in the photo conversion efficiency of the solar cells. XRD and impedance curve measurement denotes enhanced crystallinity and decrease in the ohmic resistance of the electron transport layer. Thus higher efficiency for the optimally exposed solar cells may be due to removal of the organic contaminants as indicated by the increase in the oxygen vacancies (XPS results) or due to lowering of surface roughness as confirmed by the AFM results, leading to improved wettability and increase in the adsorption of dye to the $TiO_2$ surface. However, further research is required to clearly identify the inherent mechanism. Moreover, additional experiments can be performed to explore the effect of UV-$O_3$ exposure on the photocatalytic activity, electrical conductivity and the recombination reaction at the dye – $TiO_2$ interface; leading to better optimized parameters for the development of efficient solar cells.

### ACKNOWLEDGEMENTS

This work was supported by the Department of Science and Technology, New Delhi, India through the projects DST/TSG/PT/2009/23 and IGSTC/MPG/PG(PKI)/2011A/48. G Lab Innovations (P) Ltd. Centre for Energy, Department of Chemistry, Department of Physics, and Central Instruments Facility IIT Guwahati are acknowledged for providing various instrument facilities.

**Notes:** The authors declare no competing financial interest.